\documentclass[twocolumn,showpacs,amsmath,amssymb,prb]{revtex4}

\usepackage{graphicx,color}
\usepackage{bm}

\begin{document}
\title{Superconducting proximity effect in interacting double-dot systems}

\author{James Eldridge$^1$, Marco G. Pala$^2$, Michele Governale$^{1}$, and J\"urgen K\"onig$^3$}
 \affiliation{
$^1$School of Chemical and Physical Sciences 
and MacDiarmid Institute for Advanced Materials and Nanotechnology, Victoria University of Wellington, PO Box 600, Wellington 6140, New Zealand\\
$^2$IMEP-LAHC (UMR 5130), Grenoble INP, Minatec, BP 257, 38016 Grenoble, France\\
$^3$Theoretische Physik, Universit\"at Duisburg-Essen and CeNIDE, 47048 Duisburg, Germany
}

\date{\today}

\begin{abstract}
We study subgap transport from a superconductor through a double quantum dot with large on-site Coulomb repulsion to two normal leads. 
Non-local superconducting correlations in the double dot are induced by the proximity to the superconducting lead, detectable in non-local Andreev transport that splits Cooper pairs in locally separated, spin-entangled electrons.
We find that the $I$--$V$ characteristics are strongly asymmetric: for a large bias voltage of certain polarity, transport is blocked by populating the double dot with states whose spin symmetry is incompatible with the superconductor.
Furthermore, by tuning gate voltages one has access to splitting of the Andreev excitation energies, which is visible in the differential conductance.
\end{abstract}

\pacs{73.23.Hk, 74.45.+c,03.67.Bg}
\maketitle

\section{Introduction}
\label{s1}

Advances in nanofabrication techniques allow nowadays the realization of hybrid nanoscale structures made up of quantum dots tunnel coupled to normal and superconducting leads,%
\cite{buitelaar02,buitelaar03,cleuziou06,jarillo-herrero06,jorgensen06,van_dam06,sand-jespersen07,eichler07,buizert07,eichler09,hofstetter09,hermann10,deacon_prb,deacon10}
which provides the possibility to study superconducting correlations in quantum-confined systems with few interacting degrees of freedom in the presence of non-equilibrium.
Subgap transport in these systems is sustained by Andreev reflection,\cite{degennes63,andreev64} a process consisting of the transfer of two electrons with opposite spin from a normal system to the superconducting condensate, or conversely, the transfer of the two electrons in a Cooper pair from the superconductor to the  normal region. 
Since Cooper pairs in a BCS superconductor are spin singlets, the latter possibility has been proposed as a way to inject entangled electrons in a normal system.\cite{lesovik01} Furthermore, if the two electrons stemming from a Cooper pair are separated spatially, for example because they end up in different leads, non-local correlations, similar to those discussed by Einstein, Podolsky and Rosen,\cite{epr} are generated. This type of non-local Andreev reflection is often referred to as Crossed Andreev Reflection (CAR).%
\cite{lesovik01,falci01,yamashita03,sanchez03,melin04,morten06,brinkmann06,kalenkov07,golubev07,kalenkov07-2} 
Andreev reflection in quantum dots 
has been the object of several theoretical studies.%
\cite{fazio98,fazio99,kang98,schwab99,clerk00,rozhkov00, cuevas01,tanaka07,pala07,karrasch08,governale08,futterer09,meng09,jonckheere09,zazunov10,stefanucci10,braggio10}

The use of double-dot systems for studying superconducting correlations in quantum dots introduces several advantages over single-dot devices.
While the (equilibrium) proximity effect is suppressed by the onsite charging energy in single-dot devices, non-local correlations between spatially separated dots are more easily achieved. For this reason, double-dot systems are ideal to study CAR, as the strong on-site Coulomb repulsion in the dots suppress the probability of two electrons entering the same dot and therefore local Andreev reflection. 
As a consequence, an appropriately biased double dot acts as a Cooper-pair splitter. 
A second advantage is the enhanced control on the Andreev excitation spectrum. In particular, a splitting of the Andreev excitation energies can be obtained by means of gate voltages only and detected by a differential-conductance measurement. 
Recently, two different versions of the double-dot Cooper pair splitter have been realized experimentally,\cite{hofstetter09,hermann10}
showing evidence for crossed Andreev reflections tunable via gate voltages.

In this paper, we study subgap transport properties of a double quantum dot tunnel-coupled to one superconducting $(S)$ and to two normal leads $(L,R)$ as shown in Fig.~\ref{setup}. 
We focus on the regime of large superconducting gap and dots' on-site Coulomb repulsion, where only non-local Andreev reflection is allowed. 
The transport characteristics strongly depend on the polarity of the applied bias voltage.
The transfer of Cooper pairs from the superconductor as entangled electrons into the normal leads occurs via a singlet state in the double dot.
When reversing the bias voltage, however, triplet states may be populated in the double dot.
Those have a spin symmetry that is incompatible with the superconducting lead, and transport is blocked.
The onset of this triplet blockade phenomenon is signaled by a pronounced negative differential conductance.

Superconducting correlations in the double dot can be described as a coherent superposition of the empty and doubly-occupied double-dot states.
The gate voltages applied to the individual dots allows for an independent control of both the detuning between empty and double occupation and the energy difference between the two states of single occupation.
The latter gives rise to a splitting in the excitation spectrum, which in turn is detectable in the differential conductance.  


The present article  is organized as  follows.  
In Section \ref{model} we introduce the model of the system and the formalism used to compute the Andreev current. 
In particular, we present an effective Hamiltonian for the double dot coupled to a superconducting lead with infinite pair potential.
In Section \ref{results} we discuss the main results of the paper: 
the triplet blockade and the splitting of the Andreev excitation energy spectrum.
Finally, in Section \ref{conclusions} we give some concluding remarks.

\section{Model and Formalism}
\label{model}

\begin{figure}
\includegraphics[width=\columnwidth]{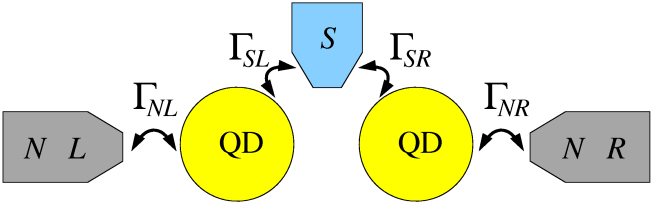}
\caption{(Color online) Schematic setup of a double quantum dot, tunnel coupled to two normal and one superconducting leads. 
The double-dot system  can be driven out of equilibrium by applying a finite bias voltages to the normal leads.}
\label{setup}
\end{figure}

The system under investigation is schematically shown in Fig.~\ref{setup}. 
It consists of a superconductor tunnel coupled to two quantum dots; each dot is contacted by a tunnel barrier with a normal metallic electrode. 
We describe the double dot by two spin-degenerate levels with interdot and intradot Coulomb-repulsion energies:

\begin{align}
\nonumber 
H_{\text{ddot}}= & \sum_{\alpha={L,R}}\left[ \varepsilon_{\alpha} \sum_\sigma d_{\alpha,\sigma}^\dagger d_{\alpha,\sigma} +U_{\alpha}
n_{\alpha,\uparrow}n_{\alpha,\downarrow}\right]+\\
& U \sum_{\sigma, \sigma'} n_{L,\sigma} n_{R,\sigma'},
\end{align}
where $\alpha=L,R$ labels the left and right dot, respectively. 
The annihilation (creation) operator for an electron in dot $\alpha$ with spin $\sigma$ is  
$d_{\alpha,\sigma}^{(\dagger)}$ and the  corresponding number operator  $n_{\alpha,\sigma}$.
The  intradot and interdot   Coulomb repulsion energies are denoted, respectively, by $U_{\alpha}$ and $U$. 

The leads are labelled by the index $\eta=L,R,S$, which indicates  the left and  right normal leads, and the superconducting lead, respectively.
The Hamiltonian of lead $\eta$ reads
\begin{equation}
H_{\eta}=\sum_{k \sigma} \epsilon_{\eta k}
c_{\eta k \sigma}^\dagger c_{\eta k \sigma}- \delta_{\eta,S}   \Delta\sum_{k}\left(  c_{\eta -k \downarrow} 
c_{\eta k \uparrow}+\text{H.c.}\right), 
\end{equation}
with the lead-electron operators $c_{\eta k \sigma}$ and $c_{\eta k \sigma}^\dagger$. 
Superconductivity in lead $S$ is described by means of the mean-field BCS Hamiltonian with a pair potential $\Delta$. 
Without loss of generality, $\Delta$ has been chosen real and positive. 
We also take the electrochemical potential of the superconductor as  reference for energies and set it to zero, $\mu_S=0$.
The tunnel coupling between the double dot and the normal leads is described by the tunneling Hamiltonian: 
\begin{equation}
H_{\text{tunn},N}=\sum_{k \sigma}\left[  V_{NL}  c^{\dagger}_{L k\sigma}d_{L,\sigma} +  V_{NR}  c^{\dagger}_{R k\sigma}d_{R,\sigma} +\mathrm{H.c.}\right]. 
 \end{equation} 
The superconducting lead is coupled to both dots and the corresponding tunneling Hamiltonian reads
\begin{equation}
H_{\text{tunn},S}=\sum_{k \sigma}\left[  V_{SL}  c^{\dagger}_{S k\sigma}d_{L,\sigma} +  V_{SR}  c^{\dagger}_{S k\sigma}d_{R,\sigma} +\mathrm{H.c.}\right]. 
\end{equation}
Finally, the Hamiltonian of the three-terminal system can be written as  $H=H_{\text{ddot}}+H_{\text{tunn},N}+H_{\text{tunn},S}+\sum_{\eta=L,R,S}H_\eta$. 

We assume that the normal-state density of states in the leads $\rho_\eta$ is constant in the energy window relevant for transport. 
We define the following tunnel-coupling strengths: $\Gamma_{NL}=2\pi \rho_L |V_{NL}|^2$ and $\Gamma_{NR}=2\pi \rho_R |V_{NR}|^2$ for the left and right normal leads, respectively, and  $\Gamma_{SL}=2\pi \rho_S |V_{SL}|^2$ and $\Gamma_{SR}=2\pi \rho_S |V_{SR}|^2$ for the coupling between the superconductor and the left and right dot, respectively.

\subsection{Large-$\Delta$ limit}
Since we are interested in transport processes involving Andreev reflection, we focus on the limit $\Delta\rightarrow \infty$, for which the superconducting lead's degrees of freedom can be integrated out exactly.
This results in the effective double-dot Hamiltonian 
\begin{align}
\nonumber
H_{\text{eff}}=& H_{\text{ddot}} - \sum_{\alpha=L,R} \frac{\Gamma_{S\alpha}}{2} \left( d^\dagger_{\alpha,\uparrow} d^\dagger_{\alpha,\downarrow}+\text{H.c}\right)+\\
& \frac{\sqrt{\Gamma_{SL}\Gamma_{SR}}}{2}( d^\dagger_{R,\uparrow} d^\dagger_{L,\downarrow}- d^\dagger_{R,\downarrow} d^\dagger_{L,\uparrow}+ \text{H. c.}) \, ,
\label{Heff}
\end{align}
which adds two different contributions to the double-dot Hamiltonian.
The first one\cite{rozhkov00} describes the local proximity effect on each dot and arises from local Andreev reflection. 
For the present context the more interesting contribution is the second term, 
which describes the formation of non-local superconducting correlations between the two dots. 
The latter is the one which encompasses Cooper pair splitting in the double-dot system.  

In realistic double-dot systems, the intradot charging energy, $U_{\alpha}$, 
amounts to several meV and is much larger than the other energy scales relevant for transport.\cite{hermann10} 
Therefore, we focus on the limit $U_{\alpha}\rightarrow\infty$, in which double occupation of each individual dot is forbidden. 
This reduces the Hilbert space of the double dot to only nine states:
the empty state $|0\rangle$, the four singly occupied states $|\alpha \sigma \rangle=d^\dagger_{\alpha,\sigma}|0\rangle$,
and four doubly occupied states with one electron in each dot. %
A convenient basis for the doubly-occupied sector of the Hilbert space is: 
the singlet $|S\rangle=\frac{1}{\sqrt{2}}\left( d^\dagger_{R,\uparrow} d^\dagger_{L,\downarrow}- d^\dagger_{R,\downarrow} d^\dagger_{L,\uparrow}\right)|0\rangle$, and the triplet states
$|T0\rangle=\frac{1}{\sqrt{2}}\left( d^\dagger_{R,\uparrow} d^\dagger_{L,\downarrow}+d^\dagger_{R,\downarrow} d^\dagger_{L,\uparrow}\right)|0\rangle$ and 
$|T\sigma\rangle=d^\dagger_{R,\sigma} d^\dagger_{L,\sigma}|0\rangle$.

In the limit of a large superconducting gap, the effective Hamiltonian Eq.~(\ref{Heff}) couples only the empty state and the singlet, 
but not the triplet states.
This coupling gives rise to the eigenstates 
\begin{equation}
|\pm\rangle= \frac{1}{\sqrt{2}}\sqrt{1\mp\frac{\delta}{2\epsilon_A}} |0\rangle \mp\frac{1}{\sqrt{2}}\sqrt{1\pm\frac{\delta}{2\epsilon_A}} |S\rangle
\end{equation}
 with energies $E_{\pm}=\delta/2\pm \epsilon_A$, where
 $\delta=\varepsilon_L+\varepsilon_R+U$ is the detuning between the empty state and the singlet. 
 The splitting between the $|+\rangle$ and the $|-\rangle$ states is  $2\epsilon_A=\sqrt{\delta^2+ 2\Gamma_S^2}$, 
where the effective coupling $\Gamma_S$ is given by 
 $\Gamma_S=\sqrt{\Gamma_{SL}\Gamma_{SR}}$.
The Andreev excitation energies are defined as the excitation energies of the double-dot system in the absence of coupling to the normal leads:
\begin{equation}
\label{abs}
E_{A, \gamma'', \gamma', \gamma} = \gamma'' \Delta \varepsilon/2+ \gamma' U/2  + \gamma \epsilon_A,
\end{equation}
with $\gamma'', \gamma', \gamma = \pm 1$, and $\Delta \varepsilon=\varepsilon_L-\varepsilon_R$.

\subsection{Master equation and current}

We integrate out the normal lead's degrees of freedom in order to obtain an effective description of the dot in terms of its reduced density matrix. 
In the limit $\Gamma_{NL,NR} \ll \epsilon_A$ and taking into account only processes in  first order in $\Gamma_{NL,NR}$, the reduced density matrix is diagonal in the eigenbasis of the effective Hamiltonian in Eq.~(\ref{Heff}) and we only need to consider the occupation probabilities $P_\chi$ of the eigenstates $|\chi\rangle$ of $H_{\text{eff}}$.
In lowest-order in $\Gamma_{NL,NR}$, the occupation probabilities obey the master equation
\begin{equation}
\sum_{\chi\ne \chi'} \left( W_{\chi \chi'}P_{\chi'}-  W_{\chi' \chi}P_{\chi}\right)=0\, ,
\end{equation}
where $W_{\chi'' \chi'}$ is the Fermi's golden rule transition rate from state $\chi'$ to state $\chi''$. 

The non-vanishing rates read
\begin{align*}
W_{\eta \sigma,\pm}=&\frac{\Gamma_{N\eta}}{2} \left( 1\mp \frac{\delta}{2\epsilon_A}\right) f^+_\eta(\varepsilon_\eta-E_\pm)\\
+& \frac{\Gamma_{N\overline{\eta}}}{4} \left( 1\pm \frac{\delta}{2\epsilon_A}\right) f^-_{\overline{\eta}}(E_\pm-\varepsilon_\eta)\\
W_{\pm,\eta \sigma}=&\frac{\Gamma_{N\eta}}{2} \left( 1\mp \frac{\delta}{2\epsilon_A}\right) f^-_\eta(\varepsilon_\eta-E_\pm)\\
+& \frac{\Gamma_{N\overline{\eta}}}{4} \left( 1\pm \frac{\delta}{2\epsilon_A}\right) f^+_{\overline{\eta}}(E_\pm-\varepsilon_\eta)\\
W_{\eta \sigma,T\sigma}=&2W_{\eta \sigma,T0}=\Gamma_{N\overline{\eta}} f^-_{\overline{\eta}}(\delta-\varepsilon_\eta)\\
W_{T\sigma\eta \sigma}=&2W_{T0,\eta \sigma}=\Gamma_{N\overline{\eta}} f^+_{\overline{\eta}}(\delta-\varepsilon_\eta)
\end{align*}
where $\eta=L,R$ here labels both the leads and the dots, and  $\overline{\eta}$ denotes the normal lead other than $\eta$, $f_\eta^{+}(\omega)$ is the Fermi function of the normal lead $\eta$ with electrochemical potential $\mu_\eta$, and $f^{-}(\omega)=1-f^{+}(\omega)$. 

The current in lead $\eta=L,R$ can be computed from the occupation probabilities by 
\begin{equation}
  I_{\eta} = \frac{e}{\hbar} \sum_{\chi\ne \chi'} 
  W_{\chi' \chi}^{ \eta} 
  P_{\chi},
\label{current}
\end{equation}
where $e<0$ is the electron charge and the current rates $W_{\chi' \chi}^{ \eta}$ take into account the number of electrons transferred to lead $\eta$.

\section{Results}
\label{results}

In this section, we present results for transport in the double-dot system in the limit of large gap and large on-site charging energy. We will consider the case of equal coupling to the normal leads, $\Gamma_{NL}=\Gamma_{NR}=\Gamma_{N}$, and of symmetric bias voltages, $\mu_L=\mu_R=\mu$. In particular we will discuss the current injected in the superconductor 
$I_S=-I_L-I_R$. 

\subsection{Triplet blockade}

\begin{figure}[h!]
\includegraphics[width=\columnwidth]{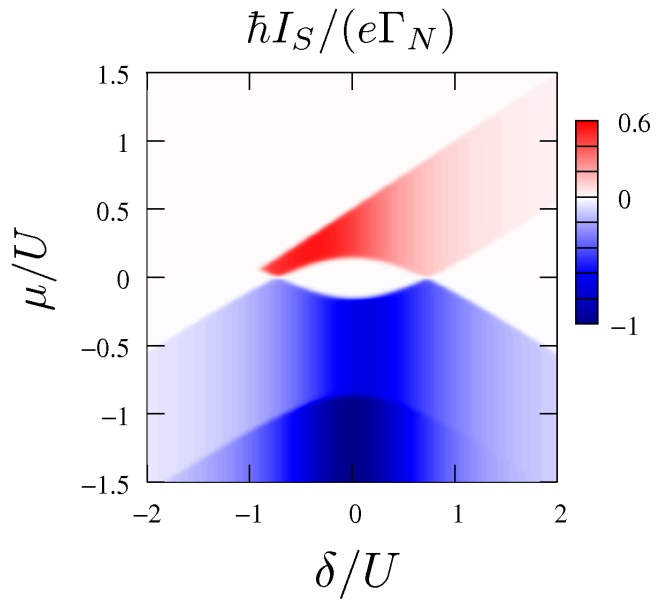}
\caption{(Color online) Density plot of the current injected into the superconducting lead $I_S$ in units of $(e \Gamma_N)/\hbar$ as a function of the detuning $\delta=\varepsilon_L+\varepsilon_R+U$ and of the chemical potential $\mu$ normal lead for degenerate dots' levels $\Delta\varepsilon=0$.
The other parameters are: $\Gamma_S=0.5 U$ and  $ k_B T=0.01 U$.  }
\label{curr_symm}
\end{figure}

\begin{figure}[h!]
\includegraphics[width=\columnwidth]{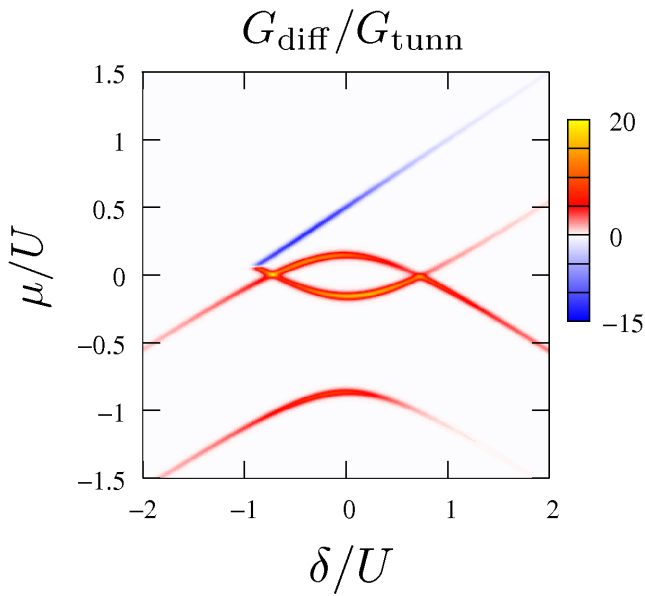}
\caption{(Color online) Density plot of the differential conductance in units of $G_{\text{tunn}}=(e^2/h) \Gamma_N/ (2 \pi   k_B T)$ as a function of the detuning $\delta=\varepsilon_L+\varepsilon_R+U$ and of the chemical potential $\mu$ of the normal leads for degenerate dots' levels $\Delta\varepsilon=0$.  The other parameters are: $\Gamma_S=0.5 U$ and  $k_B T=0.01 U$.  }
\label{diff_symm}
\end{figure}

If the sign of the applied bias voltage is such that Cooper pairs leave the superconducting electrode, then the pairs are split through the double dot and entangled electrons are emitted into the leads. For the opposite bias voltage, two electrons in the singlet double-dot state can enter the superconductor. However, if the dot is in one of the triplet states, transport is blocked and no sub-gap current can flow. Therefore, we expect the current-voltage characteristics to be asymmetric with respect to the applied voltage, with a blockaded region for voltages favoring double occupation of the double dot. 
This type of blockade is purely due to the triplet states not matching the symmetry of the pair amplitude in the BCS superconductor, and we refer to it as triplet blockade. 
In Fig.~\ref{curr_symm}, we show the current in $S$ as a function of the normal leads' chemical potential $\mu$ and of the detuning $\delta$ for the case of degenerate level positions in the two dots, $\varepsilon_L=\varepsilon_R$. The blockaded region is apparent and it is defined by $\mu>(\varepsilon_L+\varepsilon_R)/2+U=(\delta+U)/2$.
At the onset of the triplet blockade, the differential conductance, $G_{\text{diff}}=e \, dI_S/d\mu$, becomes negative, see Fig.~\ref{diff_symm}.
The triplet blockade is not lifted by removing the degeneracy of the dot levels, since this will not introduce any coupling between the triplet states and the singlet.
In the $\mu<0 $ region of Fig. ~\ref{curr_symm}, Cooper-pair splitting occurs with a strong a peak around zero detuning due to strong proximity effect on the dot. 
The peaks of positive differential conductance in Fig.~\ref{diff_symm}  map the dispersion as a function of detuning of the Andreev excitation energies. 
Since the limit of infinite on-site repulsion has been considered, it is obvious that transport is only due to crossed Andreev reflection and that pairs are split with unitary efficiency. 

\subsection{Andreev excitation energy splitting}

\begin{figure}[h!]
\includegraphics[width=\columnwidth]{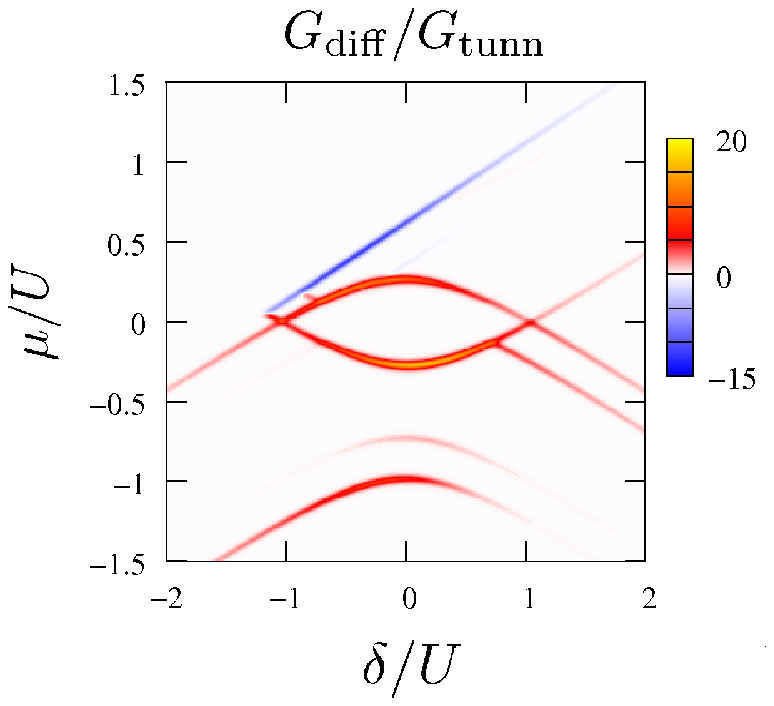}
\caption{(Color online) Density plot of the differential conductance in units of $G_{\text{tunn}}=(e^2/h) \Gamma_N/ (2 \pi   k_B T)$ as a function of the detuning $\delta=\varepsilon_L+\varepsilon_R+U$ and of the chemical potential $\mu$ of the normal leads for non-degenerate dots' levels $\Delta\varepsilon=0.25 U$.  The other parameters are: $\Gamma_S=0.5 U$ and  $k_B T=0.01 U$.  }
\label{gdiff}
\end{figure}

In the case that the dot level positions are non degenerate, $\Delta\varepsilon\ne0$, the Andreev excitation energies are split as described by Eq.~(\ref{abs}).
This splitting is controlled only by gate voltages, but it is analogous to the Zeeman splitting of the Andreev excitation energies in a single dot. 
The splitting of the excitation energies appears in the resonances in the differential conductance of the system, as can be seen in Fig.~\ref{gdiff}, where the differential conductance  is plotted as a function of the chemical potential $\mu$ and the detuning $\delta$ for a fixed value of $\Delta\varepsilon \ne 0$. 
The large negative differential conductance pinpoints the boundary of the blockaded region. 
It has to be noticed, that even if the left-right symmetry is broken due to the finite value of $\Delta\varepsilon$, 
still the two currents flowing in the normal leads are the same, i.e. $I_L=I_R$.

\section{Conclusions} \label{conclusions}
We have studied subgap transport from a superconductor through a double quantum dot with large on-site Coulomb repulsion to two normal leads.
Non-local superconducting correlations in the double dot give rise to non-local Andreev transport splitting Cooper pairs in locally separated, spin-entangled electrons.
When the polarity and magnitude of the applied bias voltage is such that electrons entering from the normal leads populate a triplet state in the double dot, then the double dot is effectively decoupled from the superconductor, and transport is blocked.
This triplet blockade shows up as an asymmetry of the $I$--$V$ characteristics. 
By  independently changing the level positions of the two dots, a splitting of the Andreev excitation energy spectrum can be achieved. This splitting can be experimentally observed in differential-conductance measurements.   

\begin{acknowledgments}
We acknowledge financial support from DFG via SFB 491.
\end{acknowledgments}


\begin{thebibliography}{50}

\bibitem{buitelaar02}
M. R. Buitelaar, T. Nussbaumer, and C. Sch\"onenberger, Phys. Rev. Lett. {\bf 89}, 256801 (2002).

\bibitem{buitelaar03}
M. R. Buitelaar, W. Belzig, T. Nussbaumer, B. Babi\'c, C. Bruder, and C. Sch\"onenberger,
 Phys. Rev. Lett. {\bf 91},057005 (2003).

\bibitem{cleuziou06}
J.-P. Cleuziou, W. Wernsdorfer, V. Bouchiat, T. Ondar\c{c}uhu, and M. Monthioux,
Nature Nanotech. {\bf 1}, 53 (2006).

\bibitem{jarillo-herrero06}
P. Jarillo-Herrero, J. A. van Dam, and L. P. Kouwenhoven, 
Nature {\bf 439}, 953 (2006).

\bibitem{jorgensen06}
H. I J{\o}rgensen, K. Grove-Rasmussen, T. Novotn\'y, K. Flensberg, and P. E. Lindelof, Phys. Rev. Lett. {\bf 96}, 207003 (2006).

\bibitem{van_dam06}
J. A. van Dam, Y. V. Nazarov, E. P. A. M. Bakkers, S. De Franceschi, and  L. P. Kouwenhoven,
Nature {\bf 442}, 667 (2006).


\bibitem{eichler07}
 A. Eichler, M. Weiss, S. Oberholzer, C. Sch\"onenberger, A. Levy Yeyati, J. C. Cuevas,  A. Mart\'in-Rodero,
 Phys. Rev. Lett. {\bf 99},126602 (2007).

\bibitem{sand-jespersen07}
T. Sand-Jespersen, J. Paaske, B. M. Andersen, K. Grove-Rasmussen, H. I. J{\o}rgensen, M. Aagesen, C. B. S{\o}rensen, P. E. Lindelof, K. Flensberg, and J. Nyg{\aa}rd,
Phys. Rev. Lett. {\bf 99}, 126603 (2007).


\bibitem{buizert07}
C. Buizert, A. Oiwa, K. Shibata, K.  Hirakawa,  and S Tarucha,  
Phys. Rev. Lett. {\bf 99}, 136806 (2007). 

\bibitem{eichler09} 
A. Eichler, R. Deblock, M. Weiss, C. Karrasch, V. Meden, C. Sch\"onenberger, and  H. Bouchiat,
Phys. Rev. B {\bf 79}, 161407(R) (2009). 

\bibitem{hofstetter09} 
L. Hofstetter, S. Csonka, J. Nyg{\aa}rd, and C. Sch\"onenberger,
Nature {\bf 461}, 960 ( 2009).

\bibitem{hermann10} 
L. G. Herrmann,  F. Portier, P. Roche, A. L. Yeyati, T. Kontos, and C. Strunk,
Phys. Rev. Lett. {\bf 104}, 026801 (2010).

\bibitem{deacon_prb}
R. S. Deacon, Y. Tanaka, A. Oiwa, R. Sakano, K. Yoshida, K. Shibata, K. Hirakawa, and S. Tarucha,
Phys. Rev. B {\bf 81}, 121308 (2010).

\bibitem{deacon10}
R. S. Deacon, Y. Tanaka, A. Oiwa, R. Sakano, K. Yoshida, K. Shibata, K. Hirakawa, and S. Tarucha,
Phys. Rev. Lett. {\bf 104}, 076805 (2010).

\bibitem{degennes63} de Gennes and D. Saint-James, Phys. Lett. {\bf 4}, 151, (1963).
\bibitem{andreev64} A.~F.~Andreev, Sov. Phys. JETP {\bf 19}, 1228 (1964). 

\bibitem{epr} A. Einstein, B. Podolsky, and N. Rosen, 
Phys. Rev. {\bf 47}, 777 (1935).


\bibitem{lesovik01}
G.B. Lesovik, T. Martin, and G. Blatter, Eur. Phys. J. B {\bf 24}, 287 (2001).

\bibitem{falci01}
G. Falci, D. Feinberg, and F.~W.~J. Hekking, Europhys. Lett. {\bf 54}, 255 (2001).

\bibitem{yamashita03}
T. Yamashita, S. Takahashi, and S. Maekawa, Phys. Rev. B {\bf 68}, 174504 (2003).
\bibitem{sanchez03}
D. S\'anchez, R. L\'opez, P. Samuelsson, and M. B\"uttiker, Phys. Rev. B {\bf 68}, 214501 (2003).
\bibitem{melin04}
R. M\'elin, and D. Feinberg, Phys. Rev. B {\bf 70}, 174509 (2004).
\bibitem{morten06}
J. P. Morten, A. Brataas, and W. Belzig, Phys. Rev. B {\bf 74}, 214510 (2006).
\bibitem{brinkmann06}
A. Brinkman, and A.A. Golubov, Phys. Rev. B {\bf 74}, 214512 (2006).
\bibitem{kalenkov07}
M. S. Kalenkov, and A. D. Zaikin, Phys. Rev. B {\bf 75}, 172503 (2007).
\bibitem{golubev07}
D.S. Golubev, and A.D. Zaikin, Phys. Rev. B {\bf 76}, 184510 (2007).
\bibitem{kalenkov07-2}
M. S. Kalenkov, and A. D. Zaikin, Phys. Rev. B {\bf 76}, 224506 (2007).


\bibitem{fazio98}
R. Fazio and R. Raimondi, Phys. Rev. Lett. {\bf 80}, 2913 (1998).

\bibitem{fazio99}
R. Fazio and R. Raimondi, Phys. Rev. Lett. {\bf 82}, 4950 (1999).

\bibitem{kang98}
K. Kang, Phys. Rev. B {\bf 58}, 9641 (1998).

\bibitem{schwab99}
P. Schwab and R. Raimondi, Phys. Rev. B {\bf 59}, 1637 (1999).

\bibitem{clerk00}
A. A. Clerk, V. Ambegaokar, and S. Hershfield, Phys. Rev. B {\bf 61}, 3555 (2000).

\bibitem{cuevas01}
J. C. Cuevas, A. Levy Yeyati, and A. Mart\'in-Rodero, Phys. Rev. B {\bf 63}, 094515 (2001).

\bibitem{rozhkov00}
A. V. Rozhkov and D. P. Arovas, Phys. Rev. B {\bf 62}, 6687 (2000).

\bibitem{tanaka07} Y. Tanaka, A. Oguri, and A.C. Hewson, New J. Phys. {\bf 9}, 115 (2007).

\bibitem{pala07} 
M. G. Pala, M. Governale, and J. K\"onig, New. J. Phys. {\bf 9}, 278 (2007).

\bibitem{karrasch08} 
C. Karrasch, A. Oguri, and V. Meden, Phys. Rev. B {\bf 77}, 024517 (2008).

\bibitem{governale08} 
M. Governale, M. G. Pala, and J. K\"onig, Phys. Rev. B {\bf 77}, 134513 (2008).

\bibitem{futterer09} 
D. Futterer,  M. Governale, M. G. Pala, and J. K\"onig, Phys. Rev. B {\bf 79}, 054505 (2009).

\bibitem{meng09} 
T. Meng, S. Florens, and P. Simon, Phys. Rev. B {\bf 79}, 224521 (2009).

\bibitem{jonckheere09}  
T. Jonckheere, A. Zazunov, K. V. Bayandin, V. Shumeiko, and T. Martin, Phys. Rev. B {\bf 80}, 184510 (2009).

\bibitem{zazunov10} 
A. Zazunov, A. L. Yeyati, and R. Egger, Phys. Rev. B {\bf 81}, 012502 (2010).

\bibitem{stefanucci10} 
G. Stefanucci, E. Perfetto, and M. Cini, Phys. Rev. B {\bf 81}, 115446 (2010).

\bibitem{braggio10}
A. Braggio, M.~Governale, M.~G.~Pala, and J.~K\"onig, arXiv:1002.4629 (2010).



\end{thebibliography}
\end{document}